\documentclass[12pt, preprint]{aastex}

\def\sqig{$\sim$}
\def\sun{$_\odot$}

\def\Reba{Bandyopadhyay}

\def\source{GX13+1}
\def\src{GX13+1}

\begin{document}
\title{
Long Term X-ray Variability in GX13+1: Energy Dependent Periodic Modulation}

\author{Robin H.D. Corbet\altaffilmark{1}}

\affil{Laboratory for High Energy Astrophysics, Code 662,\\
NASA/Goddard Space Flight Center, Greenbelt, MD 20771}
\altaffiltext{1}{Universities Space Research Association}
\email{corbet@gsfc.nasa.gov}

\begin{abstract}
A search is made for periodic modulation in the X-ray flux from the low mass
X-ray binary GX13+1 using Rossi X-ray Timing Explorer All Sky Monitor data
collected over a period of almost seven years.  From a filtered data set,
which excludes measurements with exceptionally large error bars and so
maximizes signal to noise, modulation is found at a period
of 24.065 $\pm$ 0.018  days.  The modulation is most clearly detectable at
high energies (5 - 12 keV). Spectral changes
are revealed as a modulation in hardness ratio on the 24 day period
and there is a phase shift between the modulation in the 5 - 12 keV
energy band compared to the 1.5 - 5 keV band.
The high-energy spectrum of GX13+1 is unusual in displaying both emission
and absorption iron line features and it is speculated that 
the peculiar spectral and timing properties may be connected.

\end{abstract}

\section{Introduction}

The low mass X-ray binary GX13+1 is a bright persistent source which
exhibits X-ray bursts (Fleischman 1985, Matsuba et al. 1995). Counterparts
have been identified in the infra-red (Naylor, Charles \& Longmore 1991,
Garcia et al. 1992) and radio (Grindlay \& Seaquist 1986) wavebands.
The X-ray and radio fluxes of \src\ are not correlated (Garcia et
al. 1988).  The IR counterpart has been seen to vary on timescales
of days to tens of days but no definite orbital period has been found
(e.g. Charles \& Naylor 1992, Groot et al. 1996, \Reba\ et al. 2002).
However Groot et al. (1996) did find maximum power at a period of 12.6
days from observations spanning 18 days.  From IR spectroscopy \Reba\
et al. (1999) derive a spectral type of K5III for the mass donating
star. This classification implies a mass of 5M\sun\ (Allen 1973) and so the mass-donor
is the primary star.

\src\ is usually classified as an ``atoll'' source but has some
characteristics such as the properties of its quasi-periodic oscillations
(QPOs)
which make it more similar to a ``Z'' source (Homan et al. 1998 and
references therein). Schnerr et al. (2003) find that while \src\ follows
a track in an X-ray color-color diagram on timescales of hours
apparently similar to
atoll sources, the count rate and power spectrum change in ways
unlike any other atoll or Z source. 
X-ray observations of \src\ with CCD detectors have
revealed the presence of a number of features that are attributed
to iron (Ueda et al. 2001, Sidoli et al. 2002). These include an
emission line near 6.4 keV, an absorption line at 7.0 keV and
a deep absorption edge at 8.83 keV. Previously the only other X-ray
binaries which had shown such iron absorption lines were the ``superluminal''
black hole candidates GRO J1655-40 (Ueda et al. 1998,
Yamaoka et al. 2001) and GRS 1915+105 (Kotani et al. 2000, Lee et al. 2002).

From one year of observations with the All Sky Monitor (ASM) on board
the Rossi X-ray Timing Explorer (RXTE) a modulation of the soft (1.3 -
4.8 keV) flux at a period of 24.7
$\pm$ 1 day was reported (Corbet 1996; hereafter C96). Subsequent ASM observations, however, apparently failed
to find confirmation of this periodic modulation. 
\Reba\ et al. (2002) examined 5 years worth of ASM data and
concluded that while there was evidence for quasi-periodicity on a timescale
of 20-30 days this modulation was not consistently present.

Here RXTE ASM light curves of GX13+1, covering just under seven
years, are analyzed
separated into three energy bands and
different methods for dealing with variable data quality are investigated.
It is concluded that the
X-ray flux does indeed show a persistent modulation at a period of 24 days
as initially reported.

\section{Observations}
%\subsection{ASM}
The RXTE ASM (Levine et al. 1996) consists of three similar
Scanning Shadow Cameras, sensitive to X-rays in an energy band of
approximately 1.5-12 keV, which perform sets of 90 second pointed
observations (``dwells'') so as to cover \sqig80\% of the sky every
\sqig90 minutes.  
Light curves are available in three energy bands: 1.5 to 3.0 keV
(``soft''), 3.0
to 5 keV (``medium''), and 5 to 12 keV (``hard'').
The Crab produces approximately 75
counts/s in the ASM over the entire energy range. Observations
of blank field regions away from the Galactic center indicate that
background subtraction may produce a systematic uncertainty of about 0.1
counts/s (Remillard \& Levine 1997). 

Two standard ASM light curve products are routinely available - one with flux
measurements from individual dwells which preserves the 90s
time resolution, and one which gives
averages of all dwells performed during each day.  
The ASM light curve of \src\
considered here covers approximately 6.7 years (MJD 50094 to 52536).
The overall light curve of \src\ as obtained with the RXTE
ASM is shown in Figure 1. The mean flux for the entire ASM
energy range is 23 counts/s and
no long term trend is obvious from the full energy-range light curve.

In C96 it was reported that the overall flux level of \src
was anti-correlated
with the hardness ratio. However, further investigation suggests
that the major contribution to this apparent effect is gain changes in
two of the three SSCs that comprise the ASM. These instrumental effects
cause slow drifts in hardness ratios dependent on
source spectrum with only sources with spectra identical
to the Crab showing no change (Remillard private communication).
A change in the channel definitions used to define the three energy
bands
also occurred at MJD 51548.625. 
This can give discontinuous jumps in the count rates in different
energy bands again dependent on source spectrum.
These two instrumental effects can both be seen
in Figure 1. 
The approximately linear trends in the soft and medium energy bands give
exactly the apparent hardness-ratio/intensity correlation reported
in C96.  
The anti-correlation reported
in C96 is thus at least primarily an instrumental artifact.
In order to remove these trends from the lower energy bands which,
in addition to giving apparent hardness ratio changes,
result in spurious low-frequency power in a periodogram,
the low and medium energy bands were corrected by fitting two linear
trends to the light curves before and after MJD 51548 and subtracting
these. The high energy light curve does not have an obvious trend
and so no correction was needed.

%\subsection{PCA}
%The PCA is described in detail by Jahoda et al. (1996).  This detector
%consists of five, nearly identical, Proportional Counter
%Units (PCUs) sensitive to X-rays with energies between 2 - 60 keV with
%a total effective area of \sqig6500 cm$^2$. The PCUs each have a
%multi-anode xenon-filled volume, with a front propane volume which is
%primarily used for background rejection.  
%The Crab produces 13,000
%counts/s for the entire PCA across the complete energy band.  The PCA spectral %resolution at 6 keV is approximately 18\%
%and the field of view is 1\degrees\ full width half maximum (FWHM).
%PCA observations of \src\ were obtained in two separate but coordinated
%programs. One program obtained primarily
%obtained observations approximately every three
%days between 1998-05-17 to 1998-07-13. The second observations
%primarily daily between 1998-06-14 to 1998-06-20.
%The data collection modes employed, in addition to the two standard
%modes running for all RXTE observations, were:
%B\_2MS\_16A\_0\_35\_Q,
%E\_125US\_64M\_36\_1S,
%SB\_125US\_0\_13\_1S,
%and SB\_125US\_14\_249\_1S 

\section{Analysis and Results}
%\subsection{ASM}

To investigate long term flux variability it is often
convenient to use the daily averaged ASM light curves rather than 
the dwell light curves. The error on these flux measurements can
vary significantly due to, for example, the different number of dwells
covering a source from day to day. The errors on flux
measurements in individual dwells can also vary
depending on factors such as the location of the source in
each SSC's field of view, proximity to the position of the Sun, and
the number and brightness of other sources in the field of view.
When searching for periodic
modulation in faint sources, for example, it can thus
be advantageous to weight data points contributions
to a power spectrum.
In weak sources this weighting may reveal periodic modulation not
otherwise easily detectable (e.g. Corbet, Finley \& Peele 1999, Corbet
et al. 1999). 
However, this procedure
is not appropriate if the variations in source flux are significantly
larger than typical data point errors. For the full energy range daily averaged
light curve of \src\ the standard deviation of the 
data points
is 2.8 counts/s and the mean error on measurements is 0.7 counts/s.
A simple direct weighting of all data points may thus not be appropriate
for an analysis of \src.
The light curve does, however, contain some points with exceptionally
large error bars because, for example, the points were obtained with
only a very small number of dwells or the observations were obtained
when the source was at a small angular distance from the Sun.
Another simple technique was therefore considered to improve
signal to noise in period searches. While the data contributions to the power
spectrum are not weighted, some points with errors larger than
an arbitrary value are completely excluded from the calculation of the power
spectrum. How this arbitrary value is chosen is discussed below. 

A further complication in weighting the data points is that
the variation in the number of dwells per day is significantly non-random.
The number of ASM observations made per day of
\src\ was investigated and a very strong modulation at a period of
52.6 days was found (Figure 2). This periodicity is
also found in the size of the error bars on the daily averaged flux
measurements. This 52.6 day periodicity is probably linked to
the precession period of RXTE's orbit. Note that, even if unweighted
techniques are used to extract power spectra, if the dwell light curve
is used instead of the daily averaged light curve, then a similar
weighting will effectively result.

In order to compare these different techniques for searching for
periodic signals in the variable quality ASM data
the effects of weighting and screening the data were investigated. 
In Figure 3 power spectra of the light curve of \src\ 
calculated in five
different ways are shown. The techniques employed were:\\
(a) using unweighted daily averages\\
(b) weighted daily averages\\
(c) unweighted daily averages with points with large error bars
removed\\
(d) weighted dwell data\\
(e) unweighted dwell data

It can be seen that the unweighted
procedure (a) shows a peak at near 24 days that is stronger compared
to the weighted procedure (b). This peak strongly increases
in significance when the points with large error bars are excluded (c).
The power spectra obtained from the dwell data (d and e) show similar
shapes to the weighted power spectrum with the weighted dwell data (d)
almost identical to (b).

To obtain the screened unweighted power spectrum (c) data points were
removed based on the size of the error bars and the power spectrum
calculated. This technique is based on the assumption that the peak
exhibited in the unweighted power spectrum arises from a real modulation
present in the data and that the  parameters of this modulation can best
be measured by utilizing a subset of the data which gives the highest
overall signal to noise ratio.
The filtering procedure was repeated with
different data exclusion thresholds until the maximum ratio of signal
peak at \sqig24 days to average power was obtained. The distribution
of error bar sizes in shown in Figure 4 with the optimum
screening threshold of an error of 1.4 counts/s marked. The peak
of the error bar distribution is at approximately
0.4 cts/s which is considerably less than the
mean error of 0.7 cts/s noted above which reflects
the contribution of an extended tail on the distribution. It was found that only a
relatively small number of points had to be excluded (8.2\% of the original
1897 points). These excluded points
all come from the extended tail and the ``core'' of the data
is completely included. Thus, the presence of a strong peak at 24 days
is unlikely to be an artifact of this screening procedure.

The power spectra of each of the ASM energy bands were next investigated
individually. For each band  data points were again reiteratively
filtered by the size of their error bars to maximize the signal near
24 days.  The fraction of data excluded was 22\%, 15\% and 4.5\% for the
soft, medium, and hard bands respectively with associated
screening thresholds were 0.5, 0.5 and 1.0 counts/s.
For comparison the mean and distribution peak of the errors in each band were: 
0.4 and 0.19 (soft), 0.35 and 0.17 (medium), and 0.39 and 0.22 (hard).
The larger fraction of data
excluded for the softer bands might be due to the relatively larger effects
of solar contamination at low energies as the soft energy band in particular
light curve exhibits very large error bars when the \src\ is
close to the position of the Sun. However other unknown effects
might also be involved.  Figure 5 shows the resulting
power spectra in terms of relative power.  That is, the power spectra
are normalized by the mean flux in each energy band and so show relative
modulation. Figure 6 shows details of the power spectra near the 24 day
period but plotted as absolute modulation.

From Figures 5 and 6 the following conclusions can be drawn:
(i) some signal is present in all three energy bands near 24 days.
However, significant independent detection of the signal could only
be made in the hard (5 to 12 keV) and summed energy bands. (ii) The
greatest {\em relative} modulation occurs in the soft band (1.5 to 3 keV).
There are, however, other peaks in the soft power spectrum of almost comparable
size to the signal near 24 days. (iii) The greatest {\em absolute}
modulation occurs in the hard band (5 to 12 keV).

The periodic modulation in each energy band
was quantified
by fitting sine waves to the light curves
and the results of these fits are given in Table 1. Note that
the periods derived from each energy band are all consistent within
the errors with the value found from the summed energy band of
24.07 $\pm$ 0.02 days. This gives additional confidence that a real period
has been detected. This period is also consistent with the value
reported in C96 of 24.7 $\pm$ 1 days. The phase of maximum flux, however,
differs between
the soft (1.5 to 3 keV) and medium (3 to 5 keV) energy bands (which
are consistent with each other) and the hard (5 to 12 keV) band
which trails by 4.8 $\pm$ 0.7 days ($\Delta \phi = 0.20 \pm 0.03$).
This phase difference is also clearly seen directly in the folded
light curves (Figure 7).
Due, at least in part, to this energy dependent phase difference
the folded hardness ratio is also seen to be modulated on the 24 day period. 
These folded light curves are rather smooth
thus justifying the parameterization using sine wave fits.

To investigate the coherency of the modulation the width
of the peak in the power spectrum was compared to that of a transform of a sine
wave sampled with the same frequency as the actual data.
For the peak in the transform of the summed energy bands we
find a FWHM of 0.00031 day$^{-1}$ and the peak
in the periodogram of the pure sine wave has a FWHM of 0.00033 day$^{-1}$.
The width of the power spectrum peak is thus fully consistent with
coherency.

In order to further investigate the stability of the \sqig24 day period we
calculated power spectra for each energy band using a sliding box to
investigate subsets of the data. Figure 8 shows that while the
strength of the modulation is variable all data segments
for the hard and summed energy bands show some signal near 24 days. This variability, along with
the problem of determining the optimal method to calculate the power
spectra, has probably contributed to the previous difficulty in determining the
reality and properties of the modulation in GX13+1 (\Reba\ et al. 2002).

\subsection{Comparison with Other Systems}

In order to compare the properties of \src\ 
the RXTE ASM light curves of several other sources were also investigated.
These were the ``GX atoll'' sources 
(Hasinger \& van der Klis 1989) which are not known to be
pulsars or transient black hole systems: GX3+1, GX9+1 and GX9+9.
In each case power spectra were constructed from the daily averaged
light curves both with and without weighting. Some sources showed
smooth variations on long timescales caused by intrinsic variability
rather than the ASM instrumental effects noted in Section 2.
Quadratic fits were therefore subtracted from
the lightcurves before the power spectra were calculated.
The resulting power spectra are shown
in Figure 9. 
In the weighted power spectra GX3+1 shows prominent peaks at
51.3 and 54.0 days. These are both close to the sampling period
of 52.6 days for \src. Thus these peaks may be an artifact. GX9+1
shows no significant peaks in the weighted power spectrum. GX9+9
shows peaks at 51.6 and 54.6 days which are again close to the
52.6 day sampling period for \src.

In the unweighted power spectra the peaks near 52 days disappear for
both GX3+1 and GX9+9. For GX9+1 a modest peak appears in the unweighted
power spectrum at 59.5 days. 
However, when the
data screening technique was employed for this source, it was found
that exclusion of data actually reduced the relative height of the peak.
This suggests that this peak in GX9+1 may be spurious.
Thus none of the other ``GX atoll'' sources show the same type
of timing properties in the RXTE ASM as \src.

%\subsection{PCA}

\section{Discussion}

Periodic modulation in X-ray binaries is known to arise from
three types of underlying physical processes: neutron star rotation period,
binary orbital period, and super-orbital modulation by a less clear physical
mechanism (White, Nagase \& Parmar 1995). 
Since neutron star rotation can be excluded because of the length
of the period, two likely possibilities remain for \src\ of orbital or
super-orbital modulation. The modulation found for \src\ is unusual
with no directly comparable modulation found for other bright 
low mass X-ray binaries observed with the RXTE ASM.

Modulation of X-ray flux from low-mass X-ray binaries
on the orbital period is only seen
for high inclination systems. In only a few cases (e.g. EXO 0748-676,
Wolff et al. 2002 and references therein)
are eclipses seen in bright sources as the mass-donating
star occults the central X-ray source. X-ray dipping is seen for
a number of sources with moderate inclination. These dips are caused
by vertical structure in the accretion disk and are
tied to the orbital period but are irregular in their morphology
and phasing. Spectral changes are typically seen during the dips with
the spectrum becoming harder. For still lower inclination systems X-ray modulation
on the orbital period
typically cannot be seen but optical modulation may be observed (e.g.
van Paradijs \& McClintock 1995) as varying aspects of the X-ray
heated mass-donating star are observed.
In all cases modulation on the orbital period should be intrinsically coherent as
it is locked to the orbital period, but deviations from strict periodicity
can arise due to, for example, changes in the phase at which
dipping occurs.

Super-orbital modulation has been seen in several systems with the
strongest effects demonstrated by
the high mass systems SMC X-1 and LMC X-4 and the intermediate mass
system Her X-1 (e.g. Ogilvie \& Dubus 2001). For low mass systems 
periodic super-orbital modulation
appears to be rare and the best evidence for such
modulation may come from X1820-30 which has a 172 day period (e.g.
Chou \& Grindlay 2001 and references therein).
Since super-orbital modulation is likely not tied to an underlying
good clock such as orbital modulation, but may instead be caused
by a mechanism such as accretion disk precession forced by radiation
pressure (e.g. Ogilvie \& Dubus 2001 and references therein), modulation
may have low coherence.
In Cyg X-2, which initially appeared to show a ``clean'' 78
day super-orbital
periodicity (Wijnands, Kuulkers, \& Smale 1996), the modulation was later found to be more complex with
the excursion times between X-ray minima characterized as a series of integer multiples of the 9.8 binary orbital period (Boyd \& Smale 2001).

If the modulation that is observed in \src\ is orbital in origin
then a requirement is that the mass-donating star must not overfill 
the predicted Roche lobe size. \Reba\ et al. (1999) find that a
mass donor of spectral type K5III, as found from their IR spectroscopy,
would fill the Roche lobe if the orbital period is about
25 days. If this spectral type is correct then it implies
that the modulation observed would have to be orbital in nature.

The folded light curves (Figure 7) show a smooth modulation over
the entire 24 day period. This indicates that the flux is being affected
at all phases rather than the periodicity being caused by, for example, a sharp eclipse or dips restricted to a limited phase range.
Although Accretion Disk Corona (ADC) sources, where the central X-ray
source is not observed directly, can exhibit rather broad modulation
the ADC sources have low X-ray luminosities and so are unlike \src\
which is estimated to have a luminosity of 4 $-$ 6 $\times$ 10$^{37}$
(d/7 kpc)$^2$ ergs s$^{-1}$ (1 - 20 keV, Matsuba et al. 1995).

The mechanism for the periodic modulation in \src\ is unclear. However,
the ``cleaner'' modulation observed in the hard ASM energy band
may be related to
the unusual high energy features reported by Ueda et al. (2001) and
Sidoli et al. (2002). The periodic modulation might be caused by material
located in different
parts of the binary system, such as different parts of
accretion disk structure, at different energies. 
Note that Smith, Heindl, \& Swank (2002) find from monitoring observations
with the RXTE Proportional Counter Array orbital periods of 12.7 and 18.5 days for
the black hole candidates
1E 1740.7-2942 and GRS 1758-258. Thus the presence of a long orbital period
for \src\ may be another common factor, along with
the iron spectral features, between \src\ and some black hole
systems.

It is noted that Schnerr et al. (2003) interpret their unusual
timing and spectral results for \src\ as showing the 
presence of an additional source of hard variable emission.
This component could plausibly be identified with the periodic component
of the flux that is seen most clearly in the hard band with the RXTE
ASM. Schnerr et al. (2003) propose that this hard component might
come from a jet and that variability could be caused by precession
of the jet or to variations in jet activity itself. However, the periodicity
seen in the hard component suggests that precession is unlikely to
drive this variability. However, variable occultation of part of the
jet by, for example, structure in the accretion disk could account for
the variability if the system inclination is sufficiently high.

It may be that a portion of the hard flux does originate in a physically
separate region of the system such as a jet,
while softer emission comes from the inner accretion disk
and/or the surface of the neutron star. If so, this could account
for the different timing properties of \src\ found at different
energies.

\section{Conclusion}

The X-ray light curve of \src\ shows strong evidence for the
presence of a periodicity near 24 days with modulation properties
that are energy dependent. The most likely origin
for this is some type of orbital modulation if the mass donor is
a Roche-lobe filling K5III star.

Because of the unusual nature of this modulation and the somewhat non-standard
technique used to maximize the signal in the ASM data
it would be desirable to confirm the 24 day period through other
observations and determine whether the modulation is present at other
wavelengths. Unfortunately no other high-quality long term observations
appear to exist. Although \src\ was observed by the all-sky monitors on both
Ariel V and Vela 5 the sensitivity of these experiments was
significantly less than that of the RXTE ASM. The Ginga all-sky monitor,
which was more sensitive than the Ariel V and Vela 5 instruments,
did not include \src\ in the objects for which light curves were produced 
(S. Kitamoto, private communication).  Infra-red light curves have been
obtained by several groups (Charles \& Naylor 1992,
Groot et al. 1996, Wachter 1996,
\Reba\ et al. 2002).  While these
show modulations on timescales of tens of days, which could be
consistent with a 24 day period, the observations do not
cover sufficient durations to demonstrate whether the 24 day period is
also exhibited in the infra-red.

The best prospect for confirmation of the 24 day period may come
from further RXTE observations.
If the RXTE ASM continues to operate for at least a few more years then
the additional observations obtained will form a statistically
independent data set
that can be investigated for the presence of the 24 day period.

Additional constraints on models that could account for the periodic
modulation may come from an investigation of whether the high
energy spectral features also vary on the 24 day period. Extended
IR observations would be valuable as they may show
whether the 24 day period exists at these
wavelengths and so constrain the system inclination.

\acknowledgments
I thank R.A. Remillard
for useful comments on the properties of the RXTE ASM and S. Kitamoto
for information on the Ginga ASM.
\pagebreak

\pagebreak
\noindent
{\large\bf Figure Captions}

\figcaption[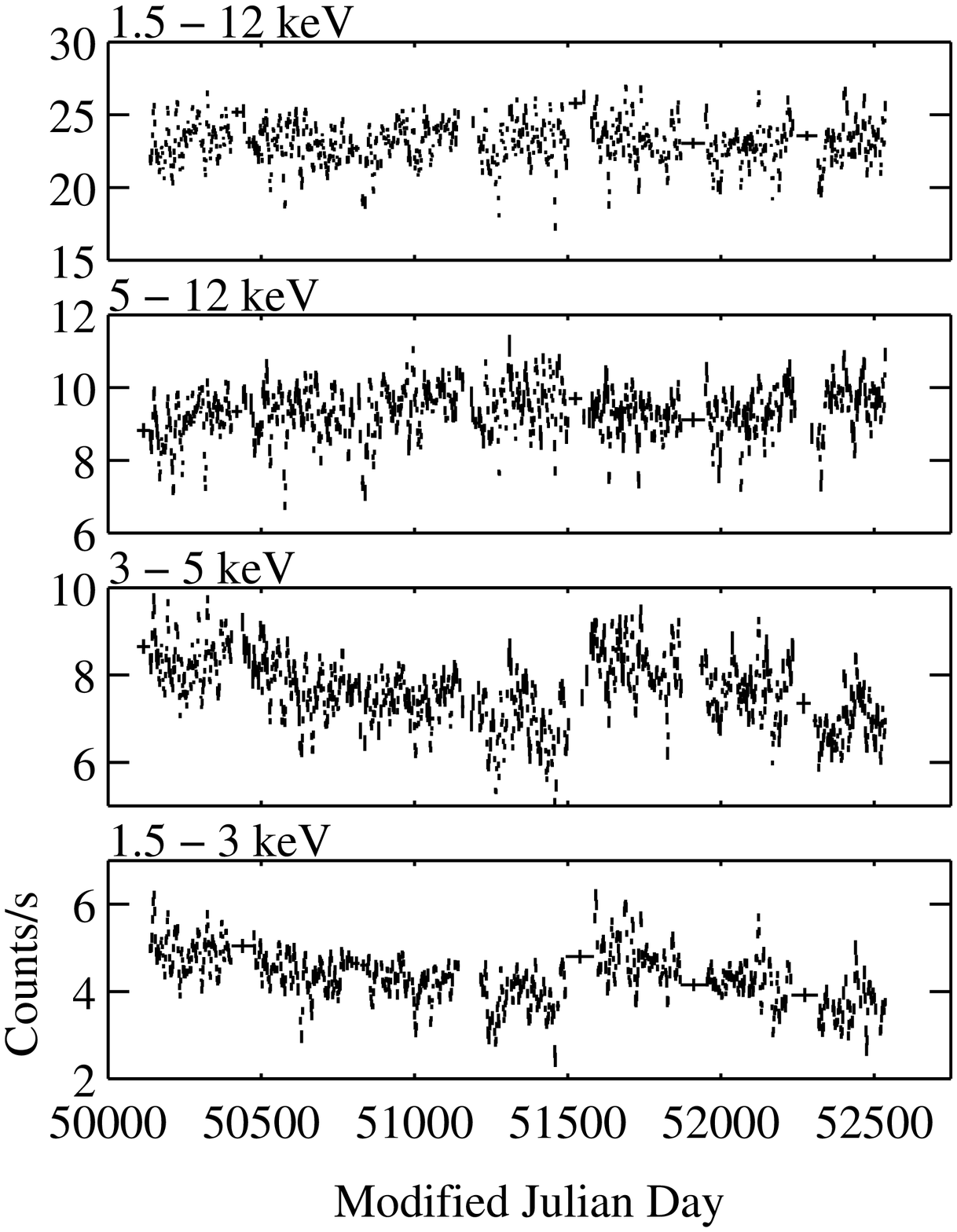]{The RXTE ASM light curve of \src\ divided into
three energy bands and the summed light curve. The linear trends
and the discontinuities in the low- and medium- energy light curves at MJD 51548.6 are due
to instrumental effects. All light curves are smoothed and rebinned versions
of the standard one day averaged light curves.}

\figcaption[f2.eps]{Lower panel: Number of ASM observations (``dwells'')
per day of \src. Upper panel: power spectrum of the curve shown in the lower panel.
The strongest power is found at a period of 52.6 days.}

\figcaption[f3.eps] {Power spectra of the ASM light curve of
\src\ obtained in five different ways. (a) Unweighted power spectrum
of daily average light curve; (b) Weighted power spectrum of daily
average light curve;
(c) Unweighted power spectrum of daily average light curve
with data filtered to exclude points
with large error bars; 
(d) Weighted power spectrum of individual dwell data;
(e) Unweighted power spectrum of  individual dwell data.}

\figcaption[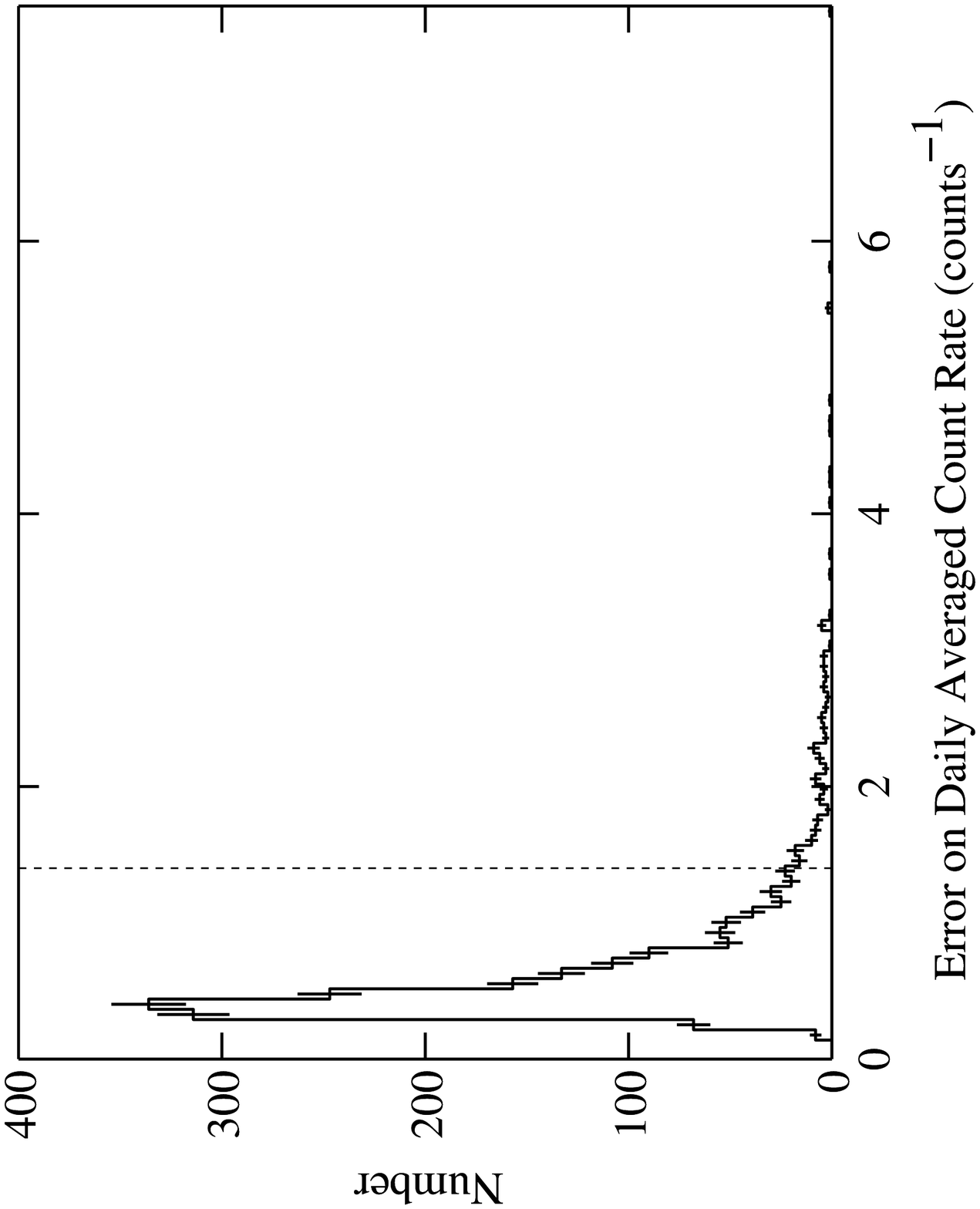]{Distribution of error bar sizes from the daily averaged
ASM light curve. The vertical dashed line shows where the cut was made
which maximizes the relative strength of the signal near
a period of 24 days in the power spectrum.}

\figcaption[f5.eps]{Power spectra of the ASM light curve of \src\ divided
into three energy bands. In each case data screening was performed to
maximize the signal at a period near 24 days. The power is normalized
to the average count rate in each energy band and is in units of ``percentage
modulation squared", i.e. the Fourier amplitude is divided by the
mean flux and the square of this quantity plotted.}

\figcaption[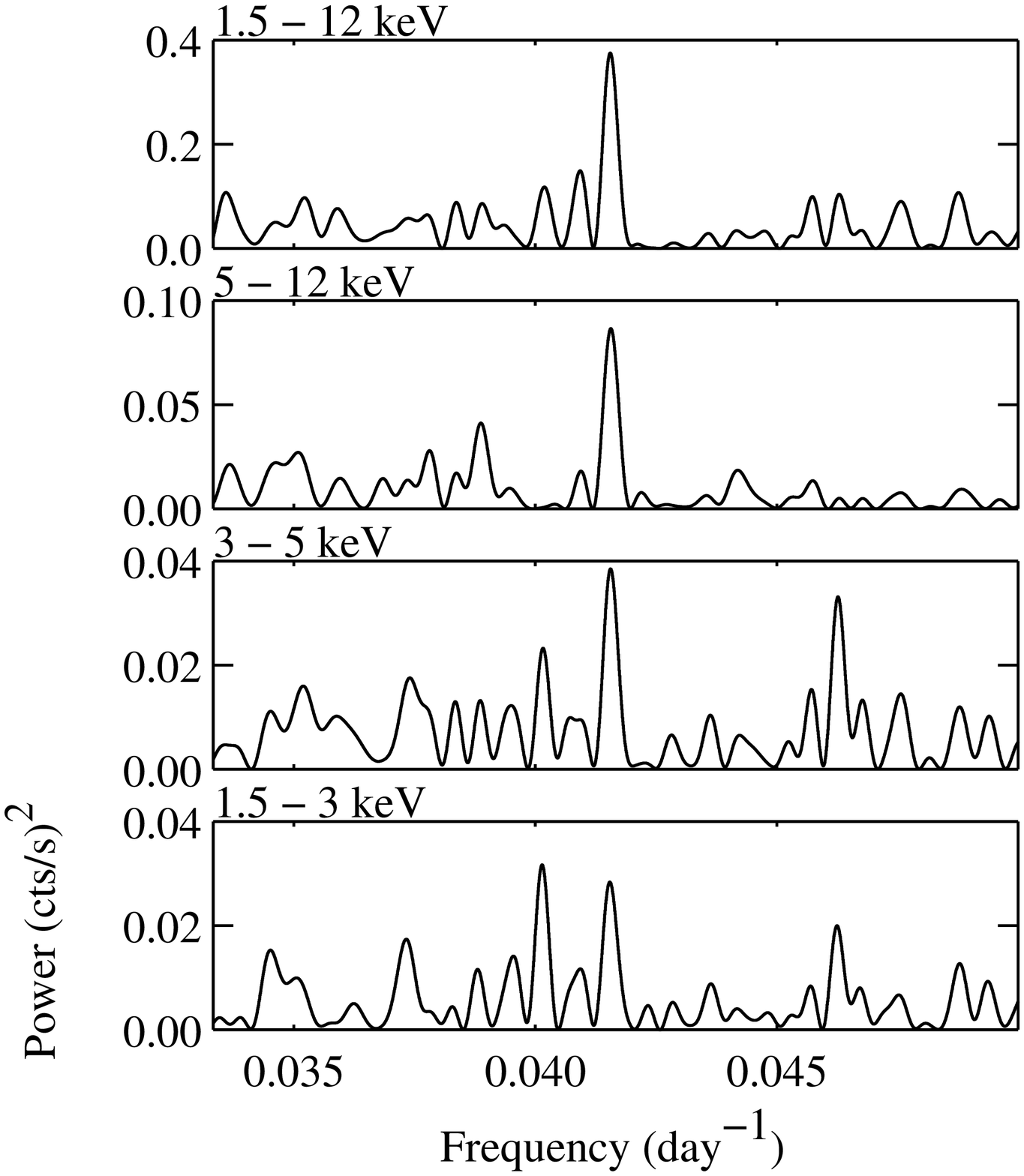]{Details of the power spectra of the ASM light
curve of \src\ near the 24 day period (equivalent to a frequency
of approximately 0.042 day$^{-1}$). For this plot the power is
{\em not} normalized to the average count rate.}

\figcaption[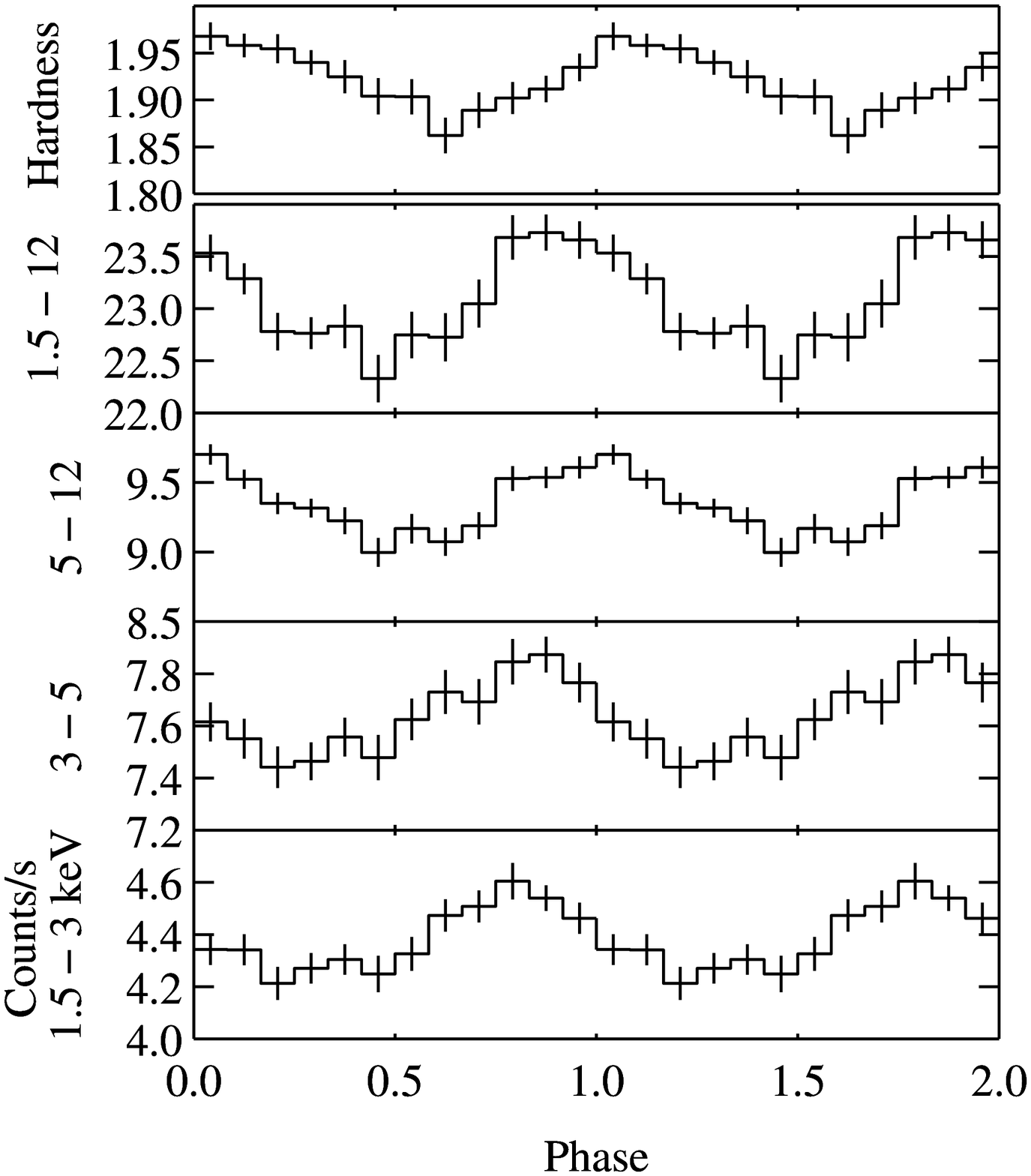]{Folded ASM light curves of \src\ and hardness
ratio (5 - 12 keV/ 1.5 - 5 keV). Two cycles are shown for clarity.
The orbital period and epoch of maximum are taken from the sine wave fit to
the hard band alone (Table 1).}

\figcaption[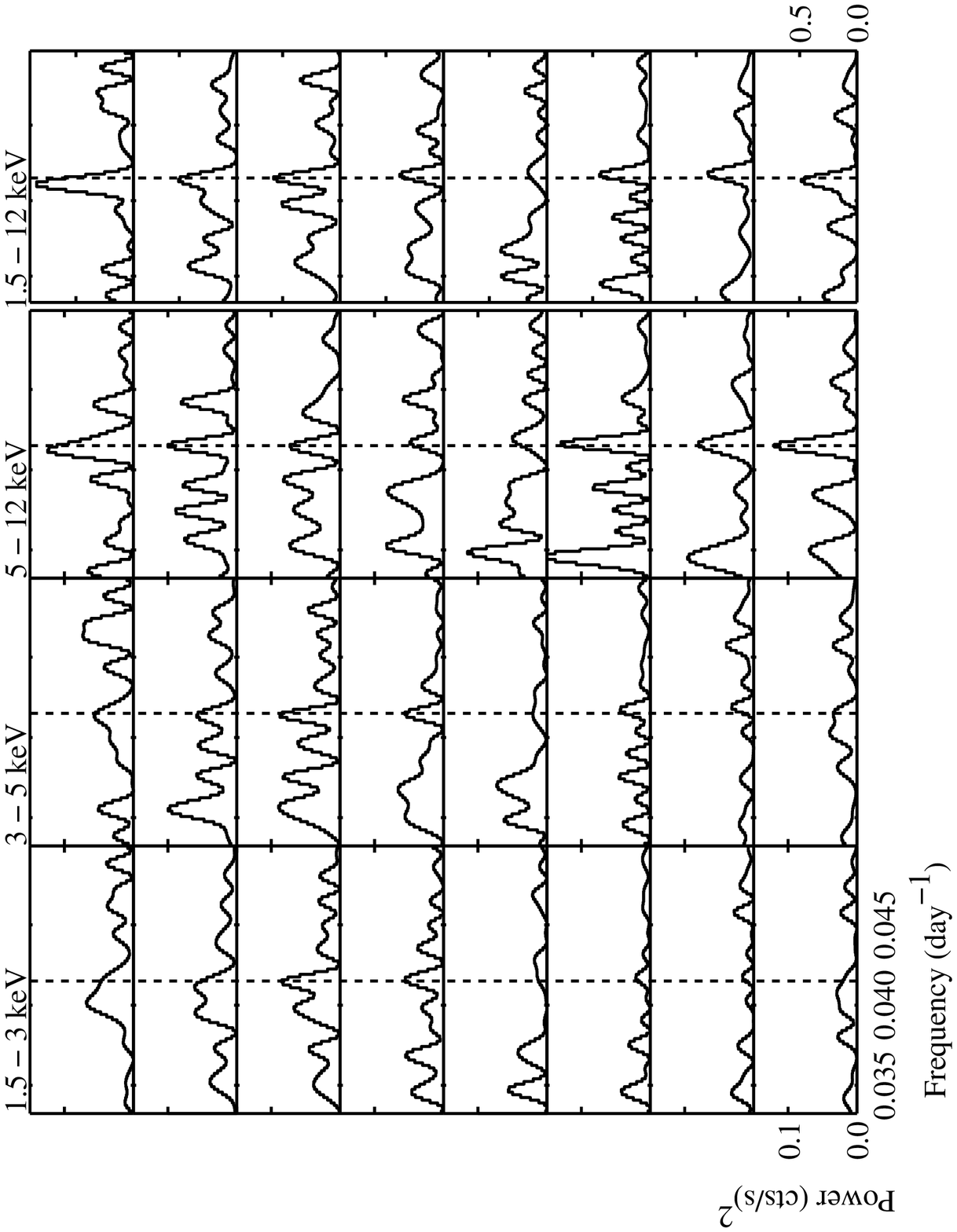]{Variations in power spectra of \src\ as a function
of time separated by energy. In each power spectrum 2 1/3 years worth of
data were analyzed with the start time of the power spectrum increased
by 6 months between each power spectrum. The individual power spectra
for each energy band are thus not statistically independent. The plots
for the energy separated power spectra are all on the same scale but the
summed energy band power spectra use a different y axis scale.  The vertical
dashed lines indicate the proposed 24.07 day period. Time intervals
correspond, from bottom panels to top, to MJDs of: 
  50137 --
  50989,
  50319 --
  51172,
  50502 --
  51354,
  50685 --
  51537,
  50867 --
  51719,
  51050 --
  51902,
  51233 --
  52085,
  51415 --
  52267
}
   
\figcaption[f9.eps]{Power spectra of RXTE ASM light curves of:
(a) GX3+1, 
(b) GX9+1, 
(c) GX9+9,
and
(d) GX13+1.
Both weighted (left) and unweighted (right) power spectra are shown.
In all cases light curves were prewhitened by
subtraction of a quadratic fit.}

\begin{table}
\caption{Sine Wave Fits to \src\ ASM Light Curves}
\begin{center}
\begin{tabular}{lccccc}
Energy Band & Mean Count Rate & Amplitude & Amplitude & Period & T$_{max}$ \\
  (keV)     & (cts/s)         &  (cts/s)  &  (\%)     & (days) & (MJD) \\
\tableline
1.5 $-$ 3   &       4.38    $\pm$ 0.01     &   0.16 $\pm$ 0.02     &     3.7     &   24.063 $\pm$ 0.019    &   51112.2 $\pm$ 0.6   \\
3 $-$ 5     &       7.64    $\pm$ 0.02     &   0.19 $\pm$ 0.03     &     2.5     &   24.062 $\pm$ 0.024      &   51112.5 $\pm$ 0.7  \\
5 $-$ 12    &       9.35 $\pm$ 0.02        &   0.30 $\pm$ 0.03   &  3.2          & 24.058 $\pm$ 0.015      &   51117.1 $\pm$ 0.5  \\
1.5 $-$ 12  &      23.10 $\pm$ 0.06        &   0.61 $\pm$ 0.08     &    2.7      &  24.065 $\pm$ 0.018     &   51115.3 $\pm$ 0.5  \\
\tableline
\end{tabular}
\end{center}
\end{table}

% Figure 1
\begin{figure}
\plotone{f1.eps}
\end{figure}

% Figure 2
\begin{figure}
\plotone{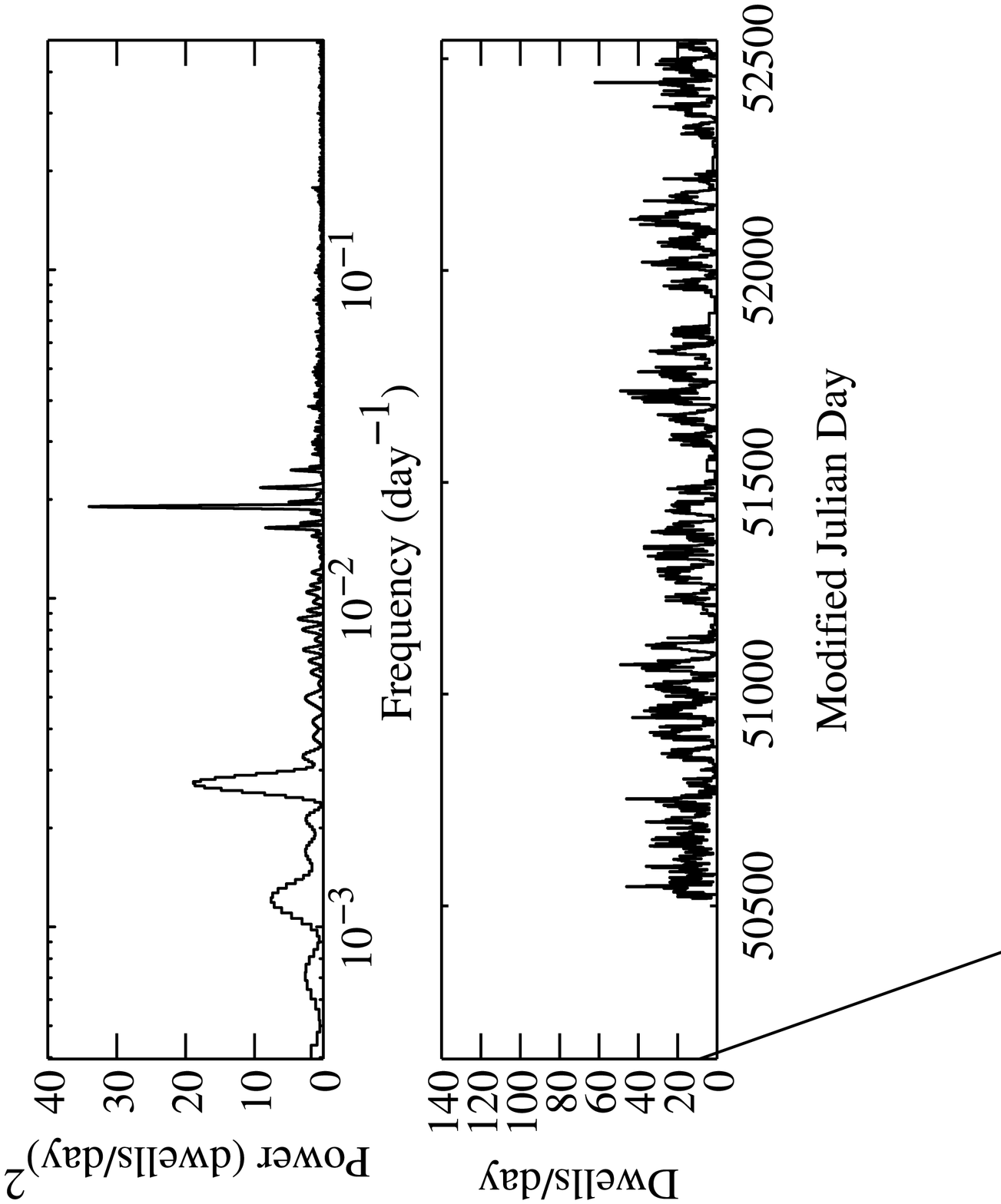}
\end{figure}

% Figure 3
\begin{figure}
\plotone{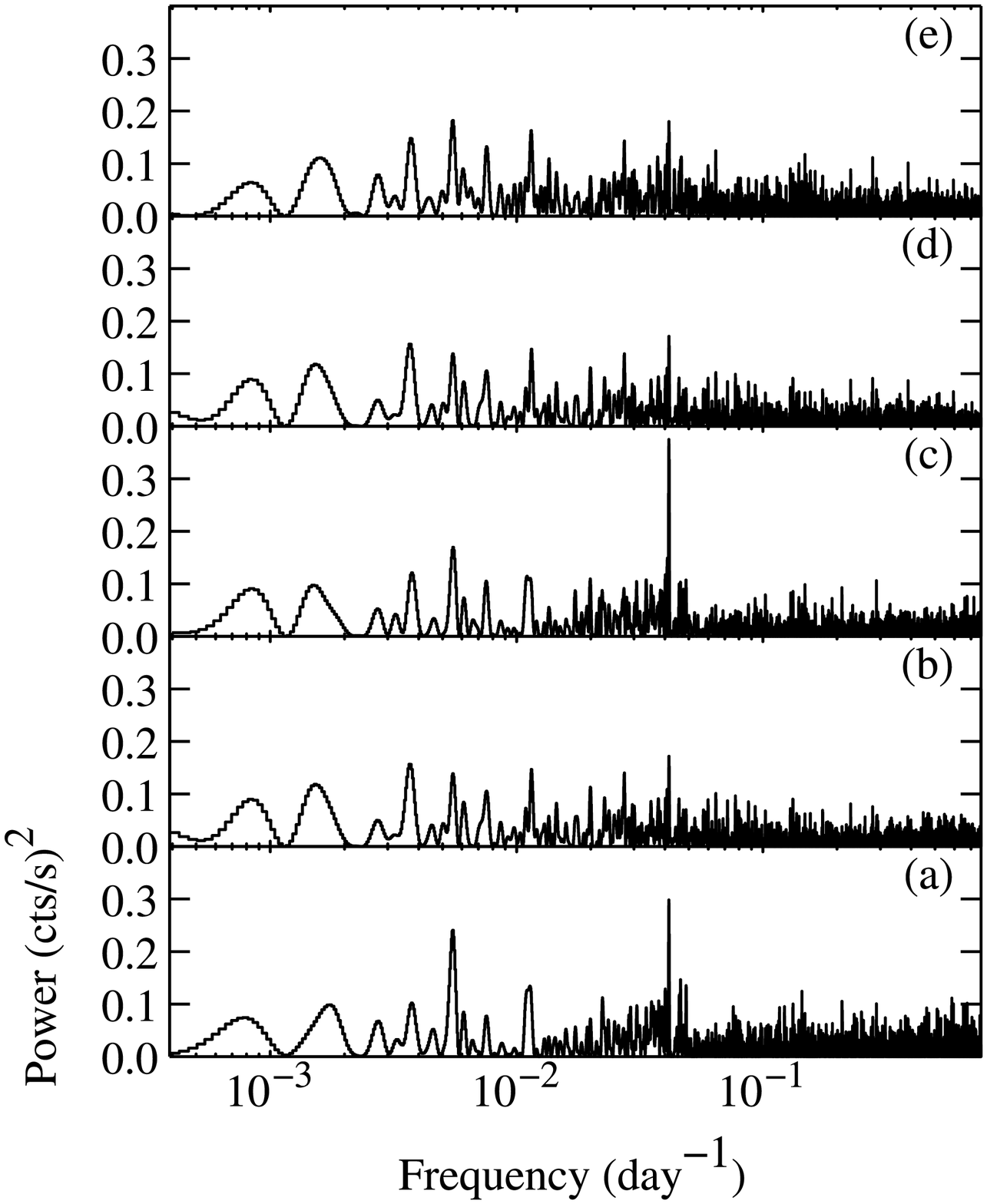}
\end{figure}

% Figure 4
\begin{figure}
\plotone{f4.eps}
\end{figure}

% Figure 5
\begin{figure}
\plotone{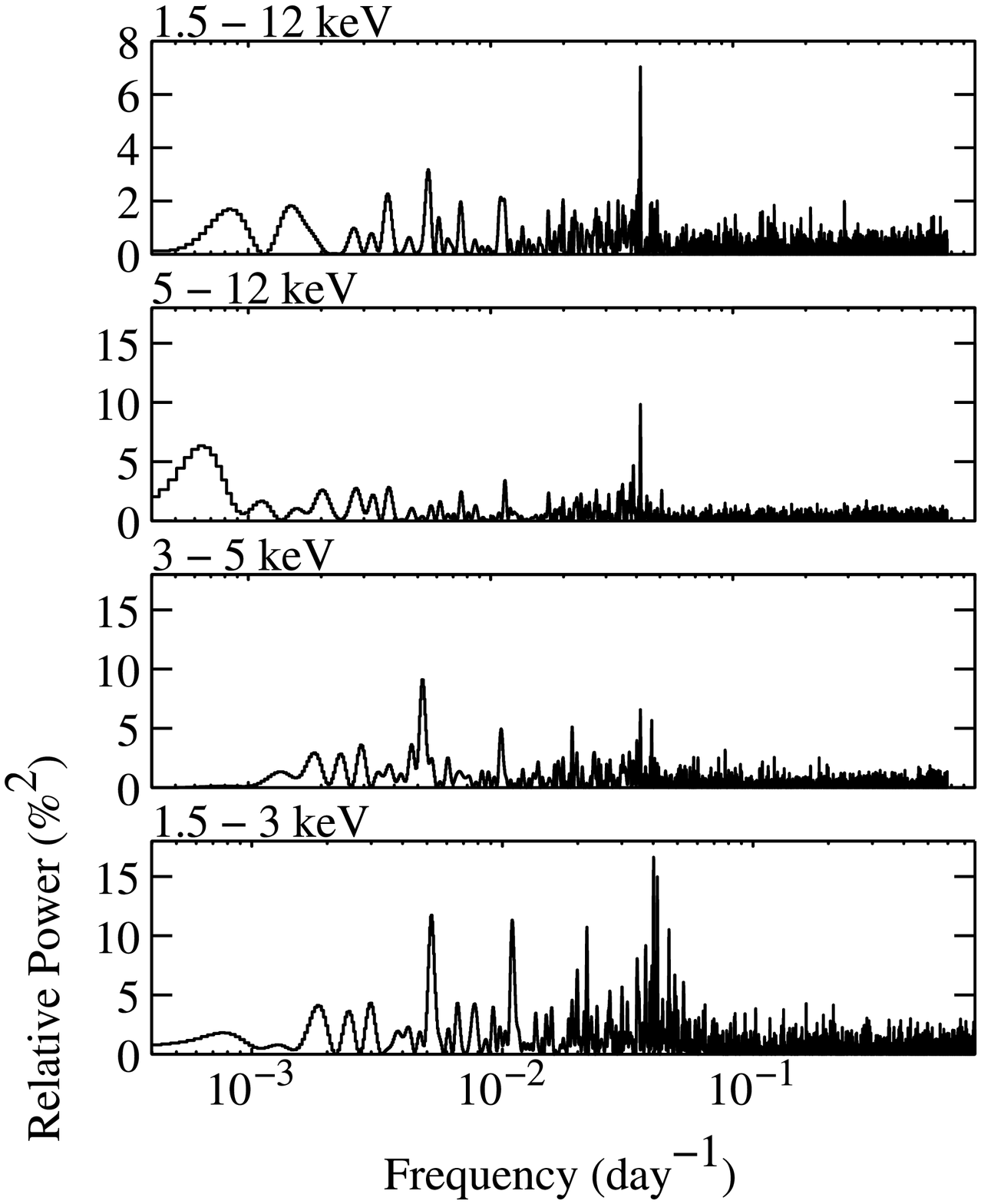}
\end{figure}

% Figure 6
\begin{figure}
\plotone{f6.eps}
\end{figure}

% Figure 7
\begin{figure}
\plotone{f7.eps}
\end{figure}

% Figure 8
\begin{figure}
\plotone{f8.eps}
\end{figure}

% Figure 9
\begin{figure}
\plotone{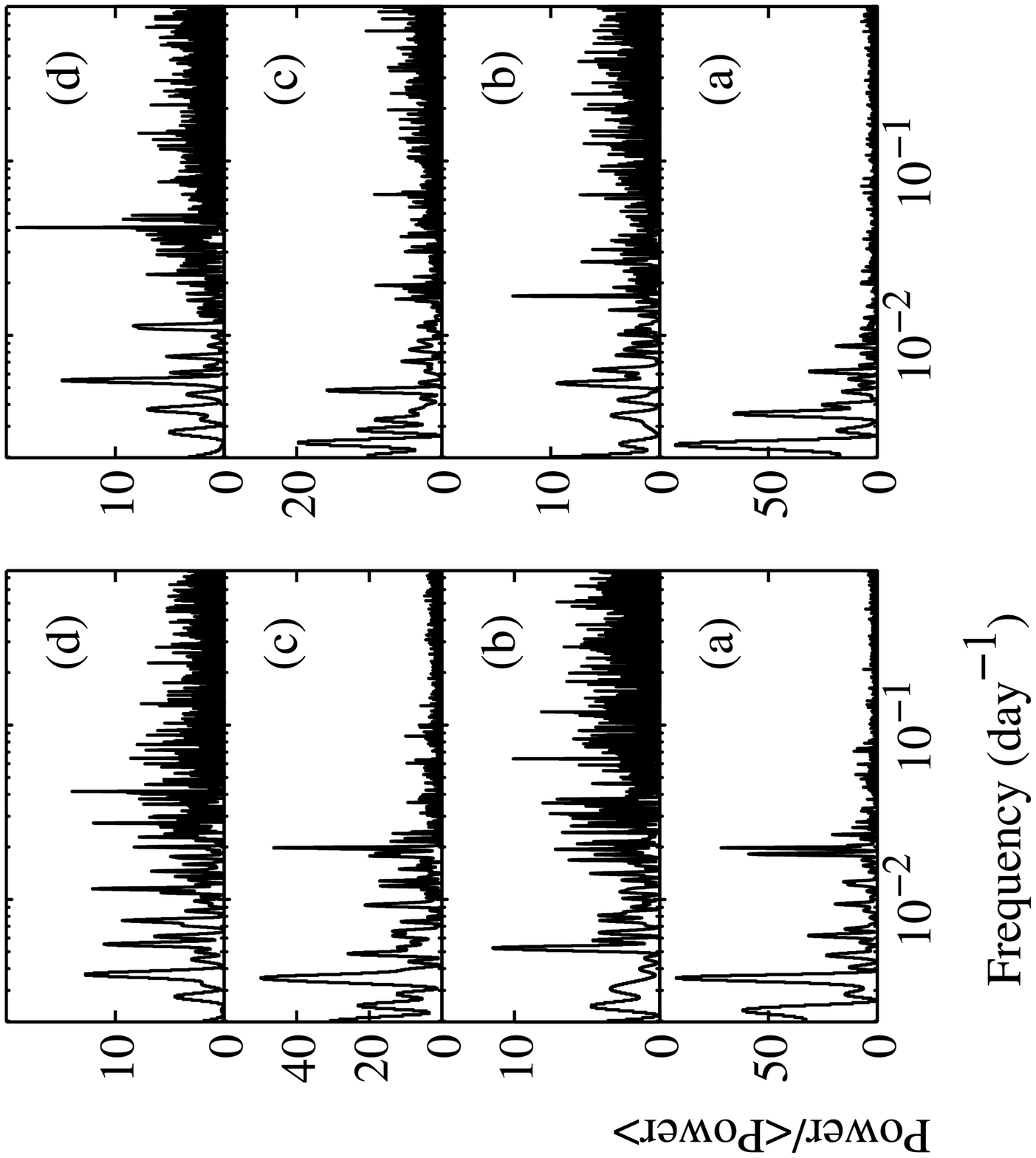}
\end{figure}


\begin{references}

\reference{}
Allen, C.W., 1973, ``Astrophysical Quantities'', Athlone Press

% IR spectroscopy
%\reference{}
%\Reba, R.M., Shahbaz, T., Charles, P. A., van Kerkwijk, M. H., 
%\& Naylor, T., 1997, \mnras, 285, 718

% IR spectroscopy II
\reference{}
Bandyopadhyay, R. M., Shahbaz, T., Charles, P. A., \& Naylor, T.,
1999, \mnras, 306, 417


%Infrared Photometric Variability of GX 13+1 and GX 17+2
\reference{}
Bandyopadhyay, R.M., Charles, P.A., Shahbaz, T., \& Wagner, R.M.,
2002, \apj, 570, 793

\reference{}
Bradt, H.V., Rothschild, R.E., \& Swank, J.H.,
1993, \aaps, 97, 355

% Discovery of IR counterpart
\reference{}
Charles, P.A., \& Naylor, T., 1992, \mnras, 225, 6P

%1820-30 variability
Chou, Y. \& Grindlay, J.E., 2001, \apj, 563, 934

\reference{}
Corbet, R., 1996, IAUC 6508 (C96)

%0114 periods
Corbet, R.H.D., Finley, J.P., \& Peele, A.G., 1999, \apj, 511, 876

% 1855 period
\reference{}
Corbet, R.H.D., Marshall, F.E., Peele, A.G., \& Takeshima, T.,
1999, \apj, 517, 956

\reference{}
Fleischman, J.R., 1985, \aap, 153, 106

% Simultaneous X-ray and radio
\reference{}
Garcia, M. R., Grindlay, J. E., Molnar, L. A., Stella, L., White, N. E.,
\& Seaquist, E. R., 1988, \apj, 328, 552

% IR counterpart
\reference{}
Garcia, M.R., Grindlay, J.E., Bailyn, C.D., Pipher, J.L., Shure, M.A.,
\& Woodward, C.E., 1992, \apj, 103, 1325

\reference{}
Grindaly, J.E. \& Seaquist, E.R., 1986,
\apj, 310, 172

\reference{}
Groot, P.J., van der Klis, M., van Paradijs, J.,
Augusteijn, T. \& Berger, 1996, in ``Cataclysmic Variables
and Related Objects'', ed. A. Evans \& J.H. Wood, p. 367

\reference{}
Hasinger, G. \& van der Klis, M., 1989,
\aap, 225, 79

% RXTE 57-69 Hz QPOS
\reference{}
Homan, J., van der Klis, M., Wijnands, R., Vaughan, B., \& Kuulkers, E.,
1998, \apjl, 499, L41  

%1915 spectrum
\reference{}
Kotani, T., Ebisawa, K., Dotani, T., Inoue, H., Nagase, F.,
Tanaka, Y., \& Ueda, Y., 2000, \apj, 539, 413

%1915 spectrum
\reference{}
Lee, J.C., Reynolds, C.S., Remillard, R., Schulz, N.S.,
Blackman, E.G., \& Fabian, A.C.,
2002, \apj, 567, 1102

\reference{}
Levine, A.M., Bradt, N., Cui, W., Jernigan, J.G.,
Morgan, E.H., Remillard, R., Shirey, R.E., \& Smith, D.A.,
1996, \apjl, 469, L33

\reference{}
Matsuba, E., Dotani, T., Mitsuda, K., Asai, K., Lewin, W.H.G., van
Paradijs, J., \& van der Klis, M., 1995, \pasj, 47, 575

% Possible counterpart of GX13+1
\reference{}
Naylor, T., Charles, P.A., \& Longmore, 1991, \mnras,
252, 203

% precessing warped disks
\reference{}
Ogilvie, G.I. \& Dubus, G., 2001, \mnras, 320, 485

% RXTE PCA obs. analyzed for spectral and power spectra changes
\reference{}
Schnerr, R.S., Reerink, T., van der Klis, M., Homan, J.,
M\'endez, M., Fender, R.P., \& Kuulkers, E.,
2003, \aap, in press.

% XMM
\reference{}
Sidoli, L., Parmar, A. N., Oosterbroek, T., \& Lumb, D., 2002,
\aap, 385, 940

% periods in 1E 1740.7-2942 and GRS 1758-258
\reference{}
Smith, D.M., Heindl, W.A., \& Swank, J.H., \apjl, 578, L129

\reference{}
Stella, L., White, N.E., \& Taylor, B.G., 1985, ESA SP-236, p. 125

% 1655
\reference{}
Ueda, Y., Inoue, H., Tanaka, Y., Ebisawa, K., Nagase, F.,
Kotani, T., \& Gehrels, N., 1998, \apj, 492, 782 (erratum 500, 1069)


% ASCA GX13+1
\reference{}
Ueda, Y., Asai, K., Yamaoka, K., Dotani, T., \&  Inoue, H., 2001,
\apjl, 556, L87

\reference{}
van Paradijs, J. \& McClintock, J.E, 1995,
in ``X-ray Binaries'', p. 58, ed. W.H.G. Lewin, J. van Paradijs,
\& E.P.J. van den Heuvel, Cambridge University Press.


\reference{}
Wachter, S., 1996, \baas, 189, 4404

\reference{}
White, N.E., Nagase, F., \& Parmar, A.N., 1995,
in ``X-ray Binaries'', p. 1, ed. W.H.G. Lewin, J. van Paradijs,
\& E.P.J. van den Heuvel, Cambridge University Press.

% claimed 78 day period in Cyg X-2
\reference{}
Wijnands, R.A.D., Kuulkers, E., \& Smale, A.P., 
1996, \apjl, 473, L45

% 0748 eclipse timing
\reference{}
Wolff, M.T., Hertz, P., Wood, K.S., Ray, P.S., 
\& Bandyopadhyay, R.M., 2002, \apj, 575, 348

% 1655 spectrum
\reference{}
Yamaoka, K., Ueda, Y., Inoue, H., Nagase, F., Ebisawa, K., Kotani, T.,
Tanaka, Y., \& Zhang, S.N., 2001, \pasj, 53, 179

\end{references}
\end{document}